\documentclass{article}

    \PassOptionsToPackage{numbers, compress}{natbib}

    \usepackage[preprint]{neurips_2020}

\usepackage[utf8]{inputenc} 
\usepackage[T1]{fontenc}    
\usepackage{hyperref}       
\usepackage{url}            
\usepackage{booktabs}       
\usepackage{amsfonts}       
\usepackage{nicefrac}       
\usepackage{microtype}      
\usepackage{graphicx}
\usepackage{amssymb}
\usepackage[ruled,vlined]{algorithm2e}
\usepackage{subcaption}
\title{Proximity Inference with Wifi Colocation during the COVID-19 Pandemic}

\author{%
  Mikhail Dmitrienko \\
  PathCheck Foundation\\
  Cambridge, MA \\
  \And
  Abhishek Singh \\
   MIT Media Lab \\
   Cambridge, MA \\
   \And
  Patrick Erichsen \\
   PathCheck Foundation \\
   Cambridge, MA \\
   \And
  Ramesh Raskar \\
   MIT Media Lab \\
   Cambridge, MA \\
}

\begin{document}

\maketitle
\begin{abstract}
In this work we propose a WiFi colocation methodology for digital contact tracing. The approach works by having a device scan and store nearby access point information to perform proximity inference. 
We make our approach resilient to different practical scenarios by configuring a device to turn into a hotspot if access points are unavailable, which makes the approach feasible in both dense urban areas and sparse rural places. We compare various shortcomings and advantages of this work over other conventional ways of doing digital contact tracing. Preliminary results indicate the feasibility of our approach for determining proximity between users, which is relevant for improving existing digital contact tracing and exposure notification implementations.
\end{abstract}
\section{Introduction}
 Privacy-preserving proximity inference is crucial to the success of a digital contact tracing solution. The majority of mainstream solutions use either Bluetooth or GPS based colocation to achieve this. However, a third option that is not as widely discussed is WiFi colocation. In this paper, we investigate a hybrid implementation of three WiFi colocation approaches described in Sapiezynski et. al, Das et al., and Carreras et. al. The hybrid implementation functions as follows: when a user is actively using WiFi, a deterministic if/else classifier infers proximity using features extracted from scans of nearby WiFi access points (APs). In the case where a user is not actively using WiFi, a duty-cycle that rotates the device between acting as a WiFi hotspot and a WiFi receiver will allow for proximity inference without interfacing with APs. We conduct an experiment to see whether three proximity features; Pearson correlation, Jaccard similarity, and Das proximity, can be used as reliable proxies for distance, and we assess the performance of a simple classifier using these features. Lastly, we assess potential privacy issues associated with WiFi-based colocation. 

\section{Related Work}

A number of groups have developed methods to perform colocalization using WiFi signals. One of the earliest papers on the subject was Meunier \cite{Meunier} in 2004, which proposed calculating the Manhattan distance between the signal strength vectors of two devices on a network to detect their proximity. That same year, Krumm \cite{krumm2004the} filed a patent for a network system in which clients send MAC address and signal strength data to a central server that estimates the likelihood of their proximity using a polynomial regression model. More contemporary papers like Sapiezynski et al \cite{sapiezynski} and Das et al \cite{das} used many of the same features from \cite{krumm2004the} to develop machine learning models. Sapiezynski et al trained a Gradient Boosting classifier while Das et al used an unsupervised, random walk algorithm which outperformed Sapiezynski's supervised approach. Other noteworthy WiFi colocation and distance estimation papers include Nakatani et al \cite{nakatani}, which trained a convolutional neural network to estimate distance for indoor navigation and Wi-Fi geo-fencing, and Carreras et al \cite{carreras}, which proposed a method that cycles devices between acting as hotspots and signal receivers to work around privacy issues associated with access point-based colocation. Following the outbreak of COVID-19, a number of groups have explored ways to leverage WiFi signal processing to help enforce social distancing guidelines and improve digital contact tracing efforts. Trivedi et al \cite{wifitrace} proposed a network-centric approach that uses maintenance logs to detect proximity between users connected to an enterprise network. A prototype of their approach has already been implemented on two college campuses. Harvard has also started using a network-centric approach for WiFi based contact tracing~\cite{harvardCT}. Gupta et al. \cite{gupta2020quest} developed a privacy preserving data collection system that allows organizations to gain insights on social distancing adoption and facilitate contact tracing using WiFi connectivity data.VContact~\cite{vcontact} proposed a method for collecting access point information similar to our method and allows infected individuals to share this information which healthy users download and query locally.

\begin{figure}
\centering
  \includegraphics[width=200pt,height=100.0312pt]{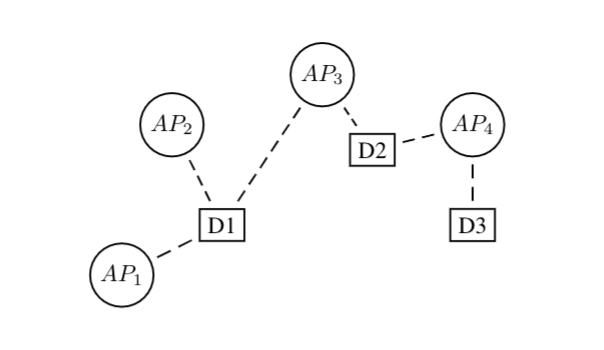}
  \caption{Devices scan for the signal strengths and MAC addresses from nearby access points. Engineered features based on this data are then used to predict proximity between users.}
  \label{fig:test1}
\end{figure}

\begin{algorithm}[b]
                \caption{Proximity classifier logic}
                Inputs: $scans_{ij}, i,j \in I, J,  avgMetrics_k$
                \\Outputs: $predictions$ \\
                
                \quad \quad $predictions \leftarrow \varnothing$ \\
                \quad \quad \textbf{for} $scan_{ij}, \forall i,j \in I,J$ \textbf{do} \\
                \quad \quad \quad \quad if $scan_{ij}[0] > avgMetrics_k[0]$ \textbf{or} $scan_{ij}[1] > avgMetrics_k[1]$ \textbf{or} \\
                \quad \quad \quad \quad $scan_{ij}[2] > avgMetrics_k[2]$ \textbf{then}\\
                \quad \quad \quad \quad \quad \quad $predictions_{ij} \leftarrow true$ \\
                \quad \quad \quad \quad \textbf{else} \\
                \quad \quad \quad \quad \quad \quad $predictions_{ij} \leftarrow false$ \\
                \quad \quad \quad \quad \textbf{end if} \\
                \quad \quad \textbf{end for}
\end{algorithm}

\section{Methods}
\subsection{Proximity Classifier} 
\label{proximityClassifier}
We propose a deterministic if/else classifier that uses the following three features: 
\begin{itemize}
    \item Pearson correlation of signal strengths from overlapping APs between two devices
    \item Jaccard similarity between the lists of APs
    \item "Proximity feature" described in Das et al \cite{das} which we will refer to as "Das proximity"
\end{itemize}

The classifier makes its prediction according to the following logic:

$scan_{ij} = \{Jaccard_{ij}, Pearson_{ij}, Das_{ij}\}$ represents the vector of the Jaccard Similarity, Pearson correlation, and Das proximity for the scan recorded by subject $i$ at distance $j$ from access point.

$avgMetrics_k = \{avgJaccard_k, avgPearson_k, avgDas_k\}$ represents the vector of the average Jaccard, Pearson, and Das at the distance threshold $k$ for proximity.
\pagebreak

                

\subsection{Hotspot Duty Cycle}
\label{headings}

In settings where an individual is not connected to a WiFi access point, we propose an implementation of a hotspot duty cycle described in \cite{carreras}. An immediate limitation of this approach is that it is only possible on Android devices - the necessary APIs are not exposed on iOS. For individuals with an Android device, Carreras et al \cite{carreras} describes four distinct advantages to an access point based approach to colocation:

i) does not require any additional access points to be present aside from the user’s device \\
\quad \quad ii) only requires information from pairwise interactions \\
\quad \quad iii) users do not need to actively connect to any access points \\
\quad \quad iv) provides accuracy in the range of 0.5m - 1.0m across a range of settings 

Native APIs are available on Android to programmatically create and destroy a hotspot on a user’s device for the duty cycle logic. In order to easily integrate this logic into a React Native app, we developed a set of React Native bindings for the underlying Android code.\footnote{Github link removed for Neurips submission.} With these bindings available, future work would include an implementation of this duty cycle in a npm package. This package would have the ability to detect when a user is connected to a WiFi access point, and if not, it would then begin to perform a hotspot duty cycle as described in \cite{carreras}. The package could be included in a React Native application to perform proximity inference with WiFi colocation while providing the least disruptive experience to an end user.

\subsubsection{Distance proxy experiment}

We launched a study to see whether the features from \ref{proximityClassifier} could be used as reliable proxies for distance. We recruited six subjects and instructed them to collect WiFi sensor log data using an Android app. Subjects recorded their first scan right next to their access point in their indoor living environment. Then, they took scans at 1 ft intervals away from their AP until they were 25ft away or as high of a distance as space permitted. By comparing the WiFi logs of the scan at each distance interval to those of the initial scan right next to the access point, we were able to capture how the three features changed with distance. These calculated features were then averaged over all the scans.

\subsection{Results}

We first assessed the viability of Pearson correlation, Jaccard similarity, and Das proximity as proxies for distance based on the scans we collected from the experiment. The change in these features averaged over all the scans can be seen in subfigures a-c. We observe that all three features steadily decreased with distance before plateauing at approximately 10ft. 

We then evaluated the performance of the classifier using standard metrics for binary classification: recall, precision, and F-score. We also found the corresponding metrics for different distance thresholds. The change in the metrics with increasing distance thresholds is captured in subfigures i-iii. Figure iii shows that F-score increased linearly with the distance threshold. Given that social distancing guidelines recommend that people remain 6-10ft apart, we aim for a high accuracy in that range. Using 10ft as the distance threshold led to an F-score of 0.65, which is comparable to the lower end of Bluetooth proximity detection algorithms described in Shankar et al \cite{shankar2020proximity}. However, due to the limited amount of data in this experiment, its difficult to quantify how well the classifier would perform in a consumer setting. We expect that further experimentation with more robust datasets will yield a more accurate picture about the feasibility of our approach.

\begin{figure}
    \begin{subfigure}[h]{0.3\textwidth}
        \includegraphics[width=130pt,height=108.0312pt]{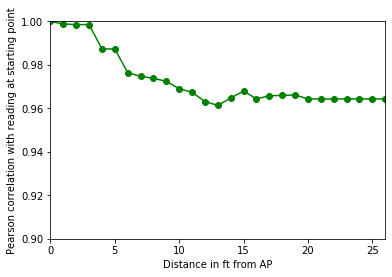}
        \caption{Average Pearson over all scans}
        \label{fig:test1}
    \end{subfigure}
    \hfill
    \begin{subfigure}[h]{0.3\textwidth}
      \includegraphics[width=130pt,height=108.0312pt]{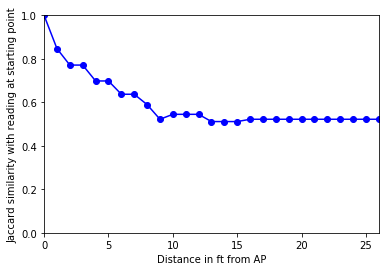}
      \caption{Average Jaccard over all scans}
      \label{fig:test2}
    \end{subfigure}
    \hfill
    \begin{subfigure}[h]{0.3\textwidth}
      \includegraphics[width=130pt,height=108.0312pt]{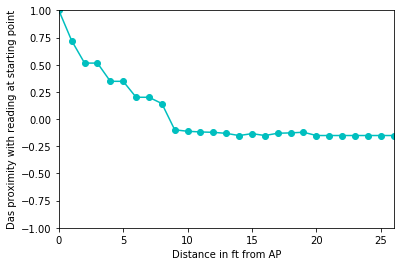}
      \caption{Average Das proximity over all scans}
    \end{subfigure}
\end{figure}

\renewcommand\thesubfigure{\roman{subfigure}}
\begin{figure}
    \begin{subfigure}[b]{0.3\textwidth}
      \centering
      \includegraphics[width=130pt,height=108.0312pt]{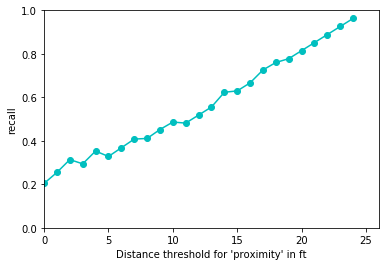}
      \caption{Change in recall with changing threshold}
      \label{fig:test1}
    \end{subfigure}
    \hfill
    \begin{subfigure}[b]{0.3\textwidth}
      \centering
      \includegraphics[width=130pt,height=108.0312pt]{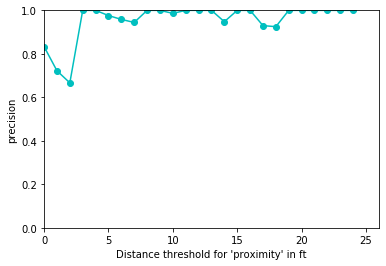}
      \caption{Change in precision with changing threshold}
      \label{fig:test2}
    \end{subfigure}
    \hfill
    \begin{subfigure}[b]{0.3\textwidth}
      \centering
      \includegraphics[width=130pt,height=108.0312pt]{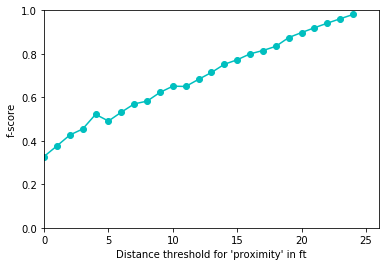}
      \caption{Change in F score with changing distance threshold}
    \end{subfigure}
\end{figure}

\section{Discussion}
\subsection{Privacy Considerations}

Privacy is one of the most critical pieces of a digital contact tracing ecosystem. Having a high level of privacy not only alleviates a majority of ethical concerns but also drives up the adoption rate, which is prerequisite for the success of any contact tracing infrastructure. While alternate solutions like Bluetooth, GPS, and ultrasound have witnessed privacy aware solutions for contact tracing in light of COVID-19, WiFi has not seen anything substantial so far. 
There are a few important differentiating factors when it comes to the privacy of a WiFi-based contact tracing solution compared to other colocation technologies. The first big difference is the lack of direct interaction between participating parties. Similar to GPS but unlike Bluetooth and ultrasound, WiFi logs its own data without exchanging any information with other participating parties. This puts a restriction on the entropy of information which can be obtained through a third party providing information, in this case a WiFi router. Therefore the only private and common information held by the two proximate parties is the MAC address. Each MAC address can only be represented by a 48 bit and hence it is not sufficient to safeguard against brute force attacks. However, we can increase the entropy by making assumptions about the additional number of hotspots/access points available which can add to the total entropy. Nevertheless, adding extra access points only leads to minimal improvements because an attacker can create a map of access points close to each other, reducing the overall entropy of the available data. The attack can be further enhanced by a dictionary mapping available through websites such as wigle.net.

\section{Conclusion}

In this proposal we outlined a two pronged approach to performing WiFi colocation during COVID-19. We demonstrated how a simple deterministic classifier could be used in a limited capacity to infer proximity using features extracted from WiFi scan data, as well as how a hotspot duty cycle could be integrated to ensure coverage of the application across places where WiFi access points may not be available. Finally, we outlined some potential privacy concerns including low information entropy and limited safeguards against brute force attacks. In the coming months, we plan on utilizing online communities to recruit more volunteers to help collect data for more nuanced experiments. Our current data set is quite limited, which prevents us from implementing more advanced machine learning algorithms without major trade-offs in test accuracy. However, deterministic classifiers for proximity detection appear to perform at comparable levels to highly sophisticated models, which presents an area for further analysis on the topic. We also hope to collaborate with other practitioners in the WiFi signal processing community to improve the proximity detecting capabilities of our particular implementation, as well as with differential privacy researchers to address the security trade offs inherent to WiFi-based contact tracing systems.

\newpage

\bibliography{refs}
\bibliographystyle{plain}

\end{document}